\documentclass[%
 reprint,
 amsmath,amssymb,
 aps,
]{revtex4-2}
\usepackage{xcolor}  
\usepackage{graphicx}
\usepackage{dcolumn}
\usepackage{bm}

\usepackage{amssymb}  
\usepackage{microtype}  
\usepackage{ragged2e}   
\usepackage[utf8]{inputenc}    
\DeclareUnicodeCharacter{2264}{$\leq$}  
\newcommand{\mnras}{Mon. Not. R. Astron. Soc.} 
\begin{document}
\newcommand{\araa}{Annu. Rev. Astron. Astrophys.} 
\newcommand{\physrep}{Phys.\ Rep.}  
\newcommand{\aap}{Astron.\ \& Astrophys.}  
\newcommand{\pasj}{Publ. Astron. Soc. Japan}  
\newcommand{\pasp}{Publ. Astron. Soc. Pac.}  
\newcommand{\apjs}{Astrophys.\ J.\ Suppl.}

\preprint{APS/123-QED}

\title{AI-Driven Reconstruction of Large-Scale Structure from Combined Photometric and Spectroscopic Surveys}

\author{Wenying Du}
 \affiliation{
 School of Physics and Astronomy, Sun Yat-sen University Zhuhai Campus, Zhuhai 519082, P. R. China.\\CSST Science Center for the Guangdong-Hong Kong-Macau Greater Bay Area, SYSU, Zhuhai 519082, P. R. China}

\author{Xiaolin Luo}%
\affiliation{%
Department of Astronomy, Shanghai Jiao Tong University, Shanghai 200240, P. R. China
}%

\author{Zhujun Jiang}%
\affiliation{%
 School of Physics and Astronomy, Sun Yat-sen University Zhuhai Campus, Zhuhai 519082, P. R. China.\\CSST Science Center for the Guangdong-Hong Kong-Macau Greater Bay Area, SYSU, Zhuhai 519082, P. R. China
}%

\author{Xu Xiao}%
 \email{xiaox87@mail2.sysu.edu.cn}
\affiliation{%
 School of Physics and Astronomy, Sun Yat-sen University Zhuhai Campus, Zhuhai 519082, P. R. China.\\CSST Science Center for the Guangdong-Hong Kong-Macau Greater Bay Area, SYSU, Zhuhai 519082, P. R. China
}%

\author{Qiufan Lin}%
\affiliation{%
Peng Cheng Laboratory, No. 2, Xingke 1st Street, Shenzhen 518000, P. R. China
}%

\author{Xin Wang}%
\affiliation{%
 School of Physics and Astronomy, Sun Yat-sen University Zhuhai Campus, Zhuhai 519082, P. R. China.\\CSST Science Center for the Guangdong-Hong Kong-Macau Greater Bay Area, SYSU, Zhuhai 519082, P. R. China
}%

\author{Yang Wang}%
\affiliation{%
Peng Cheng Laboratory, No. 2, Xingke 1st Street, Shenzhen 518000, P. R. China
}%

\author{Fenfen Yin}%
\affiliation{%
 School of Physics and Astronomy, Sun Yat-sen University Zhuhai Campus, Zhuhai 519082, P. R. China.\\CSST Science Center for the Guangdong-Hong Kong-Macau Greater Bay Area, SYSU, Zhuhai 519082, P. R. China
}%

\author{Le Zhang}%
 \email{zhangle7@mail.sysu.edu.cn}
\affiliation{%
 School of Physics and Astronomy, Sun Yat-sen University Zhuhai Campus, Zhuhai 519082, P. R. China.\\CSST Science Center for the Guangdong-Hong Kong-Macau Greater Bay Area, SYSU, Zhuhai 519082, P. R. China
}%

\author{Xiao-Dong Li}%
 \email{lixiaod25@mail.sysu.edu.cn}
\affiliation{%
 School of Physics and Astronomy, Sun Yat-sen University Zhuhai Campus, Zhuhai 519082, P. R. China.\\CSST Science Center for the Guangdong-Hong Kong-Macau Greater Bay Area, SYSU, Zhuhai 519082, P. R. China
}%

\collaboration{CLEO Collaboration}

\date{\today}

\begin{abstract}
Galaxy surveys are crucial for studying large-scale structure (LSS) and cosmology, yet they face limitations--imaging surveys provide extensive sky coverage but suffer from photo-$z$ uncertainties, while spectroscopic surveys yield precise redshifts but are sample-limited. To take advantage of both photo-$z$ and spec-$z$ data while eliminating photo-$z$ errors, we propose a deep learning framework based on a dual UNet architecture that integrates these two datasets at the field level to reconstruct the 3D photo-$z$ density field. We train the network on mock samples representative of stage-IV spectroscopic surveys, utilizing CosmicGrowth simulations with a $z=0.59$ snapshot containing $2048^3$ particles in a $(1200~h^{-1}\rm Mpc)^3$ volume. Several metrics, including correlation coefficient, MAE, MSE, PSNR, and SSIM, validate the model’s accuracy. Moreover, the reconstructed power spectrum closely matches the ground truth at small scales ($k \gtrsim 0.06~h/\rm Mpc$) within the $1\sigma$ confidence level, while the UNet model significantly improves the estimation of photo-$z$ power spectrum multipoles. This study demonstrates the potential of deep learning to enhance LSS reconstruction by using both spectroscopic and photometric data. 
\end{abstract}

\maketitle


\section{Introduction}

The most urgent cosmological questions today require the development of galaxy surveys with unprecedented depth and scale. These surveys map the large-scale structure (LSS) of the Universe, offering critical and independent constraints on cosmological models. In particular, wide and deeper imaging surveys play a vital role in accurately determining cosmological parameters.  

Imaging surveys enable the measurement of transverse Baryon Acoustic Oscillations (BAO)~\citep{Padmanabhan_etal2007,EstradaSefusattiFrieman2009,Hutsi2010,Seo_etal2012,Carnero_etal2012,deSimoni_etal2013,Abbott:2017wcz,DES:2021esc,Chan_xip2022,DESY6_BAO,Song_etal2024}. Recent transverse BAO measurements, such as those in~\cite{DESY6_BAO}, have achieved precision competitive with spectroscopic surveys. In addition, imaging surveys offer precise measurements of weak gravitational lensing~\citep{ 2003ARA&A..41..645R,Albrecht:2006um,2008PhRvD..78f3503J,2013PhR...530...87W,2015APh....63...23H,BartelmannSchneider2001,Heymans_etal2013,Hildebrandt_etal2017,Troxel_eltal2018,Asgari_etal2021,Hikage_etal2019,Amon_etal2022,Secco_etal2022,Li_etal2023,Dalal_etal2023}, a highly effective technique for probing the distribution of dark matter, the dynamics of dark energy, the formation of LSS, and the fundamental nature of gravity on cosmological scales. Specifically, cosmic shear is often employed for this purpose in combination with galaxy-galaxy lensing and galaxy clustering, forming a comprehensive approach to cosmological analysis~\citep{DESY1_3x2pt, DESY3_3x2pt, Heymans_etal2021, Miyatake_etal2023,Sugiyama_etal2023}. These studies offer robust constraints on $S_8$, achieving uncertainties of approximately 2\%, comparable to those reported by the Planck collaboration~\citep{Planck2020}.

Upcoming Stage IV surveys, including the Rubin Observatory Legacy Survey of Space and Time (LSST)~\citep{LSST_2019,LSST_DESC_2018}, Euclid~\citep{Euclid_2011}, the Chinese Space Station Telescope (CSST)~\citep{Zhan_2011,Gong_etal2019}, and the Roman Space Telescope (formerly WFIRST)~\citep{WFIRST_2015,WFIRST_2021}, are expected to provide an unprecedented wealth of photometric data, enabling even tighter cosmological constraints. It is clear that the future of cosmology relies on the development of efficient and powerful algorithms capable of effectively extracting meaningful information from the vast volume of data collected.

In contrast, spectroscopic surveys, which directly measure galaxy spectroscopic redshifts (spec-$z$) with typical precision of $10^{-4}$, enable detailed 3D clustering analyses, including anisotropic BAO and redshift-space distortions (RSD). However, these surveys are expensive and time-consuming, significantly limiting the amount of data they can collect.

However, the blurring of LSS in the radial direction due to the photometric redshift error degrades measurements of the clustering pattern~\citep{salvato2019many,brescia2021photometric, newman2022photometric}, typically yielding redshift errors of $\Delta z \sim 0.01$--0.02 and corresponding to radial distance uncertainties of approximately 40--80 Mpc. To enhance accuracy, one promising approach is to increase the number of observed color bands, as demonstrated in multi-band surveys like J-PAS~\citep{J-PAS:2014hgg}, achieving redshift errors on the order of $\Delta z = 0.003(1 + z)$.

Several widely used approaches for estimating photo-$z$ are template fitting and training methods. The template fitting method~\citep{Arnouts_1999, Bolzonella_etal2000, Benitez_2000, arnouts2002measuring,  maraston2005evolutionary, Ilbert_etal2006} involves fitting observed colors or magnitudes to synthetic models constructed from known spectral energy distribution (SED) templates, with the photo-$z$ treated as a fitting parameter. Prior information can also be incorporated into the fitting process. However, the accuracy of this method is constrained by the quality and representativeness of the SED templates, as well as the reliability of the prior information used. 

 Alternatively, clustering-based methods offer an independent approach to calibrating the true redshift distribution of a photo-$z$ sample. These methods are categorized into clustering-$z$ and self-calibration techniques, depending on the use of an external spectroscopic redshift (spec-$z$) sample. The clustering-$z$ method cross-correlates the photo-$z$ sample with an external spec-$z$ sample~\citep{Newman_2008, MatthewsNewman_2010, McQuinnWhite_2013, Menard_etal2013, Schmidt_etal2013, Morrison_etal2017, vandenBusch_etal2020, Gatti_etal2018, Gatti_etal2022, Cawthon_etal2022, Hildebrandt_etal2021, Rau_etal2023}, while the self-calibration method relies solely on the photometric sample’s clustering~\citep{Schneider:2006br, Zhang_etal2010, Benjamine_etal2010, Zhang_etal2017, Peng_etal2022, XU2023}, with a latest approach combining it with clustering-$z$~\citep{refId0}. Recently, \citep{2023arXiv230103581T} develop a method to mitigate photometric redshift uncertainties using a Bayesian forward modelling approach.

In recent years, machine learning techniques have been increasingly applied to photometric redshift estimation and calibration, including neural network~\citep{firth2003estimating, ball2004galaxy, CollisterLahav_2004, vanzella2004photometric, Sadeh:2015lsa, DeVicente:2015kyp, Zhou_etal2021, LiNapolitano_etal2022}, decision trees~\citep{suchkov2005census}, Gaussian process regression~\citep{way2006novel, foster2009stable, way2009new, bonfield2010photometric, way2011galaxy}, support vector machines~\citep{wadadekar2004estimating}, integrated modeling~\citep{way2009new}, random forests~\citep{carliles2007photometric}, and KD-trees~\citep{csabai2003application}. More recently, deep learning techniques for photo-$z$ estimation were further developed, including multilayer perceptrons (MLP)~\citep{menou2019morpho, bilicki2018photometric, vanzella2004photometric}, convolutional neural networks (CNN)~\citep{treyer2024cnn, pasquet2018deep, mu2020photometric, schuldt2021photometric}, recurrent neural networks (RNN)~\citep{escamilla2022neural}, autoencoders~\citep{frontera2019representation, eriksen2020pau},  the Contrastive Learning and Adaptive KNN~\citep{Lin:2024kzy} and generative adversarial networks (GAN)~\citep{luo2024imputation}.

An effective way to combine spectroscopic and photometric surveys is to take advantage of their respective benefits to enhance the accuracy of photometric redshifts. This can be done by performing spectroscopic observations on a selected set of galaxies to acquire accurate redshift measurements, which can then be used to refine and calibrate photometric redshift models. For example, \cite{tosone2023augmenting} suggested using graph neural networks (GNNs) to improve redshift estimation by incorporating information from spectroscopic neighboring galaxies. Furthermore, the spatial distribution of galaxies, forming a distinctive filamentary structure known as the ``cosmic web'' on scales of several hundred megaparsecs~\citep{Bardeen:1986}, provides additional constraints on galaxy positions. 
Therefore, information from the cosmic web observed in spectroscopic surveys can be used  to enhance the redshift calibration of photo-$z$ samples. Recently, the ``PhotoWeb'' method~\citep{Aragon-Calvo:2014gsa} has integrated cosmic web constraints, represented as a probability density function of the density distribution, into photometric redshift estimates. This approach reduces typical galaxy redshift errors to $\Delta z \sim 0.001$. Another technique, the stochastic order redshift (SORT) method~\citep{Tejos:2017cky, Kakos:2022ibg} refines photo-$z$ estimate by using the precise number density  distribution over redshift from a reference sample, thus improving measurements of galaxy correlation functions.

To further improve redshift estimation accuracy, conducting joint analyses of photometric and spectroscopic surveys, while maximizing the extraction of both linear and non-linear information encoded in LSS of the Universe, remains challenging.

In this paper, we propose a deep learning approach as a more comprehensive solution for integrating photo-$z$ and spec-$z$  data. Our objective is to maximize the information extracted from both data types, enabling the full utilization of these two samples to provide an accurate estimate of redshift (or equivalently, radial distance) on LSS. Specifically, we introduce a dual UNet neural network that effectively integrates photo-$z$ and spec-$z$ data at the field level, and validate its performance by comparing the reconstructed photo-$z$ field with the true-$z$ field that represents a redshift-error-free photo-$z$ field.

This paper is organized as follows. Sect.~\ref{sect:data} describes the simulated data used in this analysis. Sect.~\ref{sect:method} presents the network architecture of the dual UNet model. Sect.~\ref{sect:result} reports and discusses the results obtained with the dual UNet model. Finally, Sect.~\ref{sect:con} offers a discussion and conclusion of our findings.

\section{Data}\label{sect:data}

In this study, we train and validate the UNet models using the~\texttt{CosmicGrowth} simulations~\citep{jing2019cosmicgrowth}, which consist of a large set of high-resolution simulations generated with tens of billions of particles. Specifically, we use the $z = 0.59$ snapshot from the \texttt{WMAP\_2048\_1200} simulation for both the training and validation of our UNet model. This simulation employs $2048^3$ particles within a volume of $(1200~h/\rm Mpc)^3$ under a $\Lambda$CDM cosmology. This simulation adopts $\Lambda$CDM cosmological parameters that agree well with the WMAP measurements~\citep{hinshaw2013nine,2011ApJS..192...18K}, specifically $\Omega_b = 0.0445$, $\Omega_m = 0.2235$, $\Omega_\Lambda = 0.7320$, $h = 0.71$, $n_s = 0.968$, and $\sigma_8 = 0.83$. For each snapshot of the simulation, groups are identified through the Friends-of-Friends (FoF) algorithm, and a corresponding FoF group catalog is created. The Hierarchical Branch Tracing (HBT) method~\citep{2012MNRAS.427.2437H} is then employed to build the merger tree and subhalo catalogs.

The high resolution of this simulation ensures that the resulting subhalo catalog achieves the typical galaxy number densities observed in current photometric surveys, making it well-suited for the objectives of this study. Both the spectroscopic and photometric mock samples used to train and test our UNet model are generated from this simulation. Spectroscopic surveys typically focus on more luminous galaxies, which are associated with more massive haLoS and subhaLoS. In contrast, photometric surveys can detect a much larger number of galaxies, although they tend to be less luminous. With this concern, we construct our spectroscopic mock sample by applying a mass threshold of $M_{\rm cut} > 3.68\times 10^{12} M_{\odot}$, resulting in a number density of $\bar{n} \simeq  1.0\times 10^{-3}~(h/\rm Mpc)^{-3}$, which is comparable to that of stage-IV spectroscopic surveys. For the photometric sample, we apply the same mass threshold, but the resulting number density is significantly higher, at $\bar{n} \simeq 1.85\times 10^{-2}~(h/\rm Mpc)^{-3}$.

Due to the Doppler effect caused by the peculiar motion of galaxies, the redshift space distortions (RSD) along the line-of-sight (LoS) direction (referred to as the $Z$ direction) were included in the mock samples. The redshift space position, $\mathbf{s}$, is related to the real-space position, $\mathbf{r}$, by the following relation:
\begin{equation}
{\mathbf s} = {\mathbf r} + \frac{\mathbf{v}\cdot{\hat z}}{aH(a)}{\hat z}\,,
\end{equation}
where $\mathbf{v}$ is the peculiar velocity of the galaxy, $a$ is the scale factor, $H$ is the Hubble parameter, and the unit vector $\hat{z}$ represents the LoS direction. The redshift measurement is then adjusted by considering an additional shift in the distance through
\begin{equation}
\Delta s_{\|} = \frac{c\Delta z}{H(z)}\,,    
\end{equation}
where $\Delta z$ represents the redshift error. For simplicity, this study assumes a Gaussian-like error distribution~\citep{Gaussian}, which is sufficient for this proof-of-concept analysis. The fiducial setting for the redshift error is as follows,
\begin{equation}
\Delta z_{\rm fid} = 0.02(1+z)\,,
\end{equation} 
corresponds to $\Delta s^{\rm fid}_{\|} \simeq 70.9~h/\rm Mpc$, which is roughly an order of magnitude larger than the RSD effect. To reduce the strong boundary effects introduced by this shift, we exclude data near the edges of the box. Specifically, we remove regions where $Z \in [0, 200]~h/\rm Mpc$ and $Z \in [1000, 1200]~h/\rm Mpc$. In these regions, the distance error $\Delta s^{\rm fid}_{\|}$ is smaller than the length in the $Z$-direction by approximately 3$\sigma$.

The density fields were computed from the catalog samples, with halos assigned to a $600^3$ mesh using the Cloud-in-Cell (CIC) scheme, and a cell resolution of $2~h/\rm Mpc^3$. This resolution effectively captures the non-linear structures of the field while avoiding the complexities associated with subhalo scales, which are less directly linked to the large-scale structures of the cosmic web. This choice has been shown to provide an appropriate resolution for deep learning-based analyses of LSS~\citep{2021ApJ...913....2W,2023MNRAS.522.4748W}. The mesh field was divided into $6 \times 6 \times 4$ chunks, each with a size of $200 ~(h/\rm Mpc)^3$. The data were divided into training and test sets based on a partition along the $Z$-direction, with approximately 75\% allocated for training and 25\% for testing. Specifically, the training set spans $Z \in [200, 800]~h/\rm Mpc$, while the test set covers $Z \in [800, 1000]~h/\rm Mpc$. The training set includes $6 \times 6 \times 3$ subboxes, and the test set consists of $6 \times 6 \times 1$ subboxes.

\section{Method}\label{sect:method}

\subsection{Dual UNet-based Network}

In this study, we propose a dual UNet-based network for integrating photo-$z$ and spectroscopic-$z$ (spec-$z$) samples to reconstruct the true-$z$ field. We extend the UNet framework by incorporating a dual architecture to account for two types of mock samples.

The UNet architecture is widely employed for field reconstruction tasks\cite{wu2023ai,li2023reconstruction,xu2024comprehensive,bao2023deep,lin2023volumetriC,peng2022deep}. It extracts features through an encoder, restores resolution with a decoder, and utilizes skip connections to transfer detailed information, which enables high-precision image segmentation. Meanwhile, the dual architecture, commonly used for evaluating the similarity between input pairs, consists of two identical sub-networks that share the same weights and structure. This configuration allows the network to assess the similarity between sample pairs by calculating the distance between their feature vectors. The dual network approach has been successfully applied in various domains (e.g., facial expression recognition , image fusion, and image splicing forgery detection), where accurately determining the similarity between input pairs is essential for reliable predictions~\cite{saurav2023dual,wang2024attention,ding2023dcu,yang2019robust}.

\begin{figure*}
    \centering
    \includegraphics[width=\textwidth]{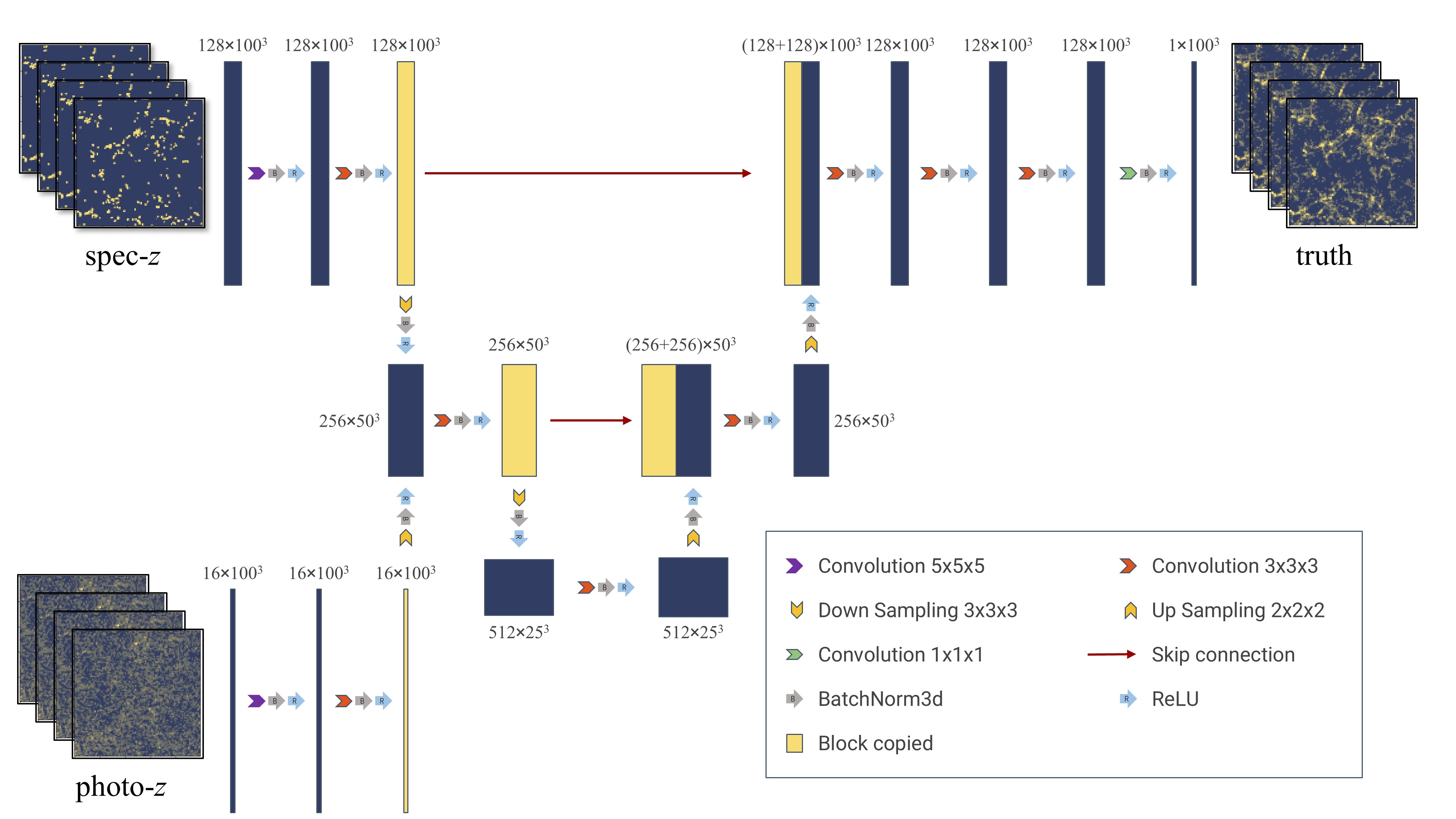}
    \caption{Architecture of the dual UNet-Based Model for the true-$z$ density field reconstruction by integrating the photo-$z$ and spec-$z$ density fields. It is trained with a learning rate of $3.9\times 10^{-4}$
  and a batch size of 2. Both the spec-$z$ and photo-$z$ density fields are fed simultaneously into the dual network, where they undergo the detailed operations illustrated in the figure. The network processes these inputs to produce the true-$z$ density field as its output. The spec-$z$ input is allocated 128 channels, while the photo-$z$ input is allocated 16 channels. Both the input and output boxes have a spatial resolution of $100^3$ pixels. This design enables the dual network to effectively integrate and extract the complementary characteristics of the spec-$z$ and photo-$z$ data for accurate reconstruction of the true-$z$ density field.}
    \label{arch-UNet}
\end{figure*}

The architecture of the dual UNet-based model we proposed is illustrated in Fig.~\ref{arch-UNet}. Specifically, the network begins by jointly processing the spectroscopic and photometric data, allocating 128 channels for the spectroscopic data and 16 channels for the photometric data. Based on our tests, increasing the number of channels did not lead to any significant improvement in reconstruction performance, suggesting that the current channel configuration is already well-optimized for the task. Each input box has a spatial resolution of $100^3$ pixels. This initial step allows the model to effectively integrate  the complementary features of the two samples for the density field reconstruction. The proposed dual UNet model employs 3D convolutional layers with kernel sizes of $5^3$ and $3^3$ to extract spatial features from the input data, capturing both local and global structures essential for accurate reconstruction. Downsampling operations using $3^3$ kernels reduce the spatial dimensions, allowing the network to learn hierarchical features at multiple scales. Conversely, upsampling with $2^3$  kernels increases the spatial resolution, enabling the network to reconstruct high-resolution outputs from lower-dimensional feature maps. The inclusion of ``skip connections'' enables the integration of low-level and high-level features. This design helps retain fine-grained details and enhances the accuracy of the reconstruction process. Copied blocks allow the network to reuse the same set of layers. Additionally, BatchNorm3d is used to stabilize training, while ReLU activation introduces non-linearity, enhancing the network's ability to model complex patterns.

For training, we set the learning rate to $3.9\times10^{-4}$ and used a batch size of 2.
The model is trained with a MSE Loss function, using gradient descent to optimize parameters and improve segmentation accuracy. To prevent overfitting, dropout layers are integrated into the architecture, effectively reducing model complexity and enhancing generalization.

\subsection{Loss Function}
In this study, we adopted the mean squared error (MSE) Loss function, which is widely used in deep learning models to assess predictive accuracy by calculating the squared difference between predicted and true values. In the context of UNet, the MSE Loss measures the discrepancy between the network's output and the target image. Specifically, it calculates the squared difference between the predicted and true values for each pixel, then averages the squared errors across all pixels. This penalizes larger prediction errors more heavily, encouraging the model to focus on minimizing them during training, ultimately improving overall prediction accuracy. The MSE Loss function is defined as:
\begin{equation}\label{eq:Loss}
    \mathcal{L} = \frac{1}{N_{\rm pix}}\sum_{i = 1}^N ({\hat{y}_i} - y_i^{\rm true})^2\,,
\end{equation}
where $N_{\rm pix}$ is the number of pixel points, $y^{\rm true}_i$ and $\hat{y}_i$ denote the true and the predicted values of the $i$-th pixel, respectively. By iteratively optimizing the MSE Loss function, UNet gradually learns more accurate image features, improving reconstruction effectiveness. 

\begin{figure}
\centering
\includegraphics[width=0.9\columnwidth]{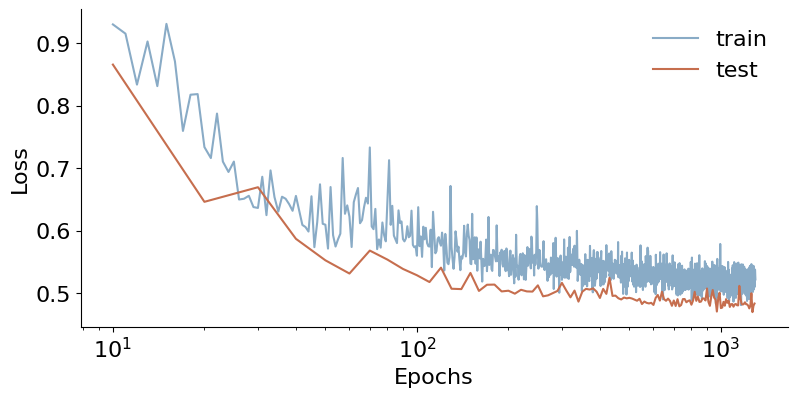}
\caption{Loss of the training (blue) and test (red) datasets. The dual UNet-based model achieves convergence after approximately 1400 epochs of training for the density field reconstruction. The Loss function utilized is the MSE Loss, as defined in Eq.~\ref{eq:Loss}.}
\label{fig:Loss}
\end{figure}

The evolution of the loss function over epochs is shown in Fig.~\ref{fig:Loss}, where the training loss (blue) and test loss (red) are plotted. The model parameters are saved upon completion of training, and the final model is selected based on its performance on the test set to mitigate overfitting. As depicted, the model converges after approximately 1400 epochs of training.

\subsection{Correction for density field on large scales}

Since the model is trained exclusively on density fields within a box size of $200^3~(h/\rm Mpc)^3$, it is not expected to accurately recover large-scale $k$-modes for $k \lesssim 2\pi/L = 0.031~h/\rm{Mpc}$. Our simulation tests confirm this, showing that the power spectrum of the dual UNet-reconstructed true-$z$ density field closely matches the ground truth for $k \gtrsim 0.031~h/\rm{Mpc}$. However, at larger scales ($k \lesssim 0.031·~h/\rm{Mpc}$), the prediction accuracy deteriorates.

To address this issue, we propose a correction for the large scales. Since the true-$z$ field is unknown in real observations, this correction relies on information from both the spec-$z$ and model-predicted power spectra at large $k$-scales, which is practical as these quantities are observable and can be derived from actual data. The correction is then applied to the small $k$-scales.

Specifically, the small $k$-modes can be corrected using a transfer function $T_f (k)$, which represents the ratio between the spec-$z$ and model-predicted 1D power spectra at $k > k_{\rm cut}$, with $k_{\rm cut}= 0.031~h/\rm Mpc$ in this study, as defined below:
\begin{equation}\label{eq:r}
T_f(k) = 
\begin{cases} 
\frac{1}{N_{k'}} \sum_{k' > k_{\rm cut}} \sqrt{\frac{P^{\rm spec}(k')}{P^{\rm model}(k')}} & \text{if } k < k_{\rm cut} \\
1 & \text{otherwise}\,,
\end{cases}
\end{equation}
where $T_f$ is calculated as the averaged ratio over $k$ bins, with a total of $N_k$ bins. $P^{\rm model}$ and $P^{\rm spec}$ represent the power spectra estimated from the large test box, which has a side length of 1200~$h/\rm{Mpc}$. The corrected density field $\delta(\mathbf{x})$ for each small test box, with a side length of 200~$h/\rm{Mpc}$, is obtained by applying $T_f$ to the corresponding Fourier modes, using the following relation: $\delta^{\rm corr}(\mathbf{x}) = \texttt{FFT}^{-1}\left\{ T_f(k) \times \texttt{FFT}\left\{\delta(\mathbf{x})\right\} \right\}$. Here $\texttt{FFT}\{\cdot\}$ and $\texttt{FFT}^{-1}\{\cdot\}$ denote the Fast Fourier Transform and its inverse, respectively.


\section{result}\label{sect:result}
This section presents the performance evaluation of the trained dual UNet-based model, based on predictions for $6\times6\times3=108$ sub-boxes from the test dataset, which were not used during model training or the optimization of its structure and parameters. Each large box has a physical size of $200~(h/\rm Mpc)^3$ and contains $100^3$ pixels.

To quantitatively evaluate reconstruction accuracy, we begin by introducing several statistical measures for subsequent analysis. One widely used metric for characterizing a density field is the two-point correlation function, expressed as
\begin{equation}
\xi_{\alpha\beta}(\mathbf{r})=\langle\delta_\alpha(\mathbf{x}) \delta_\beta(\mathbf{x}+\mathbf{r})\rangle\,,
\end{equation}
where $\alpha$ and $\beta$ denote two arbitrary fields, and $\delta(\mathbf{x})$ denotes the density contrast at a point $\mathbf{x}$. The separation vector, $\mathbf{r}$, specifies the separation between two points. The ensemble average, $\langle\cdot\rangle$, is calculated by averaging over all spatial points $\mathbf{x}$.

The correlation function, $\xi_{\alpha\beta}(\mathbf{r})$, is cLoSely related to the power spectrum, which can be categorized into the auto-power spectrum ($\alpha = \beta$) and the cross-power spectrum ($\alpha \neq \beta$). This relationship is established through the Fourier transform:  
\begin{equation}\label{eq:pk}
P_{\alpha\beta}(\mathbf{k})=\int \xi_{\alpha\beta}(\mathbf{r}) \mathrm{e}^{i \mathbf{k} \cdot \mathbf{r}} \mathrm{d}^3 \mathbf{r}\,,
\end{equation}  
where $\mathbf{k}$ is the three-dimensional wavevector of the plane wave, with magnitude $k = |\mathbf{k}|$ and a corresponding wavelength $\lambda$ given by $k = 2\pi / \lambda$.

Furthermore, for a given reconstructed density field $f$ from UNet, the correlation coefficient, $C_r$, enables a direct comparison between the reconstructed field $f$ and the true field $f'$, which is defined as:
\begin{equation}\label{eq:R}
C_r = \frac{1}{N_{\rm pix}-1}\sum_{i}\frac{ (f_i-\bar{f}) (f'_i-\bar{f'})}{\sigma_f \sigma_{f'}}\,,
\end{equation}
where $N_{\rm pix}$ denotes the total number of pixels, which is the same for both fields. The sample mean and standard deviation of field $f$ are denoted by $\bar{f}$ and $\sigma_f$, respectively. For an ideal reconstruction, where the reconstructed field perfectly matches the true field, the correlation coefficient reaches its maximum value of $C_r=1$.

In addition, the Mean Absolute Error (MAE) is another metric that quantifies the average of the absolute differences between predicted and true values. Mathematically, MAE is defined as:
\begin{equation}\label{eq:mae}
    {\rm MAE} = \frac{1}{N_{\rm pix}}\sum_{i = 1}^N \left|{\hat{y}_i} - y_i^{\rm true}\right|\,.
\end{equation}
Unlike MSE, which squares the differences, MAE provides a more direct measure of the average error magnitude. It treats all errors equally, without giving more weight to larger deviations.

Moreover, to evaluate the quality of the reconstruction, the Peak Signal-to-Noise Ratio (PSNR) is employed, defined as a log-scaled version of the Mean Squared Error (MSE):

\begin{equation} \label{eq:PSNR}
  {\rm PSNR}(x_{\rm true}, x_{\rm rec}) = 10 \log_{10}\left( \frac{L^2}{{\rm MSE}(x_{\rm true}, x_{\rm rec})}\right),
\end{equation}

where $L = |x_{\rm max} - x_{\rm min}|$ represents the dynamic range of the ground truth map. Higher PSNR values correspond to greater reconstruction accuracy.

Additionally, the Structural Similarity Index Measure (SSIM)~\citep{1284395} is used to assess the structural similarity between the true and reconstructed maps. SSIM values range from 0 to 1, with higher values indicating better performance. It is computed as:

\begin{equation} \label{eq:SSIM}
  {\rm SSIM}(x_{\rm true}, x_{\rm rec}) = \frac{(2\mu_i\mu_j + C_1)(2\Sigma_{ij} + C_2)}{(\mu^2_{i} + \mu^2_{j} + C_1)(\sigma^2_{i} + \sigma^2_{j} + C_2)},
\end{equation}

where $\mu$ and $\sigma$ denote the mean and variance of the maps, respectively, with subscripts $i$ and $j$ referring to the true and reconstructed maps. $\Sigma_{ij}$ represents their covariance. The constants $C_1 = (k_1 L)^2$ and $C_2 = (k_2 L)^2$, with $k_1 = 0.01$ and $k_2 = 0.031$ , ensure numerical stability and prevent division by zero. SSIM agrees with human visual perception, making it a robust metric for evaluating reconstruction quality.

\subsection{Visual inspection and point-wise comparison}

\begin{figure*}
    \centering
    \includegraphics[width=\textwidth]{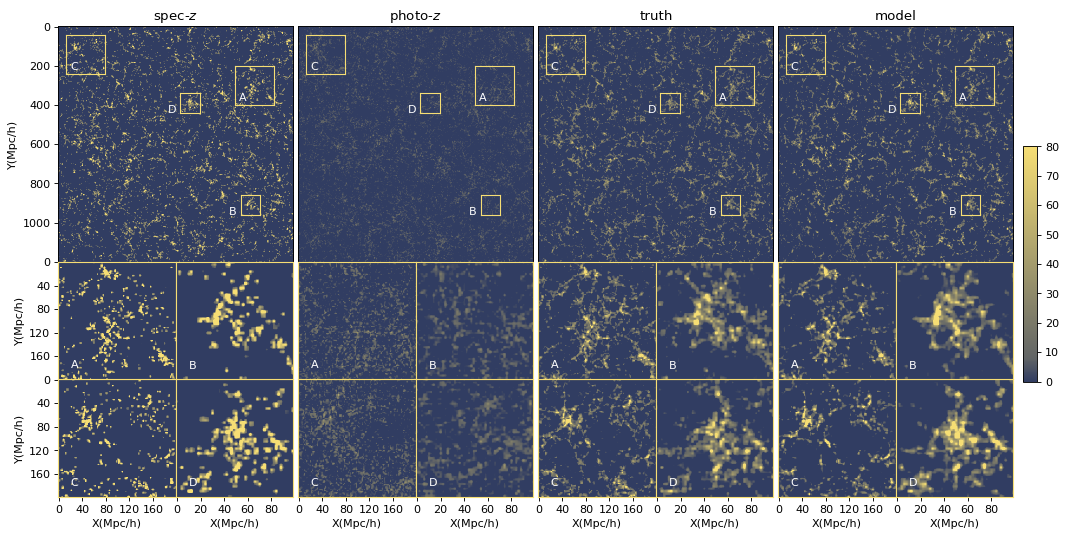}
    \caption{Comparison of the density fields for the ground truth, the reconstruction, the spec-$z$, and the photo-$z$ (from left to right). For visualization, each plot represents a slice of the density field projected along LoS with a bin width of $10~{\rm Mpc}/h$. To highlight small-scale details, four regions--labeled A, B, C, and D--are randomly selected, and their zoomed-in views are displayed in the lower panels. As demonstrated, the reconstruction performs well across both large and small scales.  }
    \label{fig:recon-field}
\end{figure*}

As illustrated in Fig.~\ref{fig:recon-field}, a comparison is made between the density fields of the ground truth, the reconstruction, the spec-$z$, and the photo-$z$ data, displayed from left to right. For a visualization purpose, each plot corresponds to a two-dimensional slice of the density field, projected along LoS with a bin width of $10~{\rm Mpc}/h$. To better emphasize small-scale features, four distinct regions--labeled A, B, C, and D--are randomly selected, and their zoomed-in views are provided in the lower panels. 

The results demonstrate that the reconstruction captures the density field with high fidelity across both large and small scales. Notably, both the sparse spec-$z$ sample and the photo-$z$ sample exhibit significant deviations from the true density field. In contrast, the dual UNet model successfully reconstructs the true field by effectively learning from a combination of these two samples. However, subtle discrepancies between the reconstructed field and the ground truth can be observed, particularly in regions with high-density contrasts and voids. These differences are more pronounced in areas where spec-$z$ samples are absent, as highlighted in the zoomed-in views.

\begin{table}
    \centering
    \caption{Reconstruction Performance for the dual UNet model we have trained. The average values of MSE, MAE, SSIM and PSNR and associated 1$\sigma$ statistical errors are shown, respectively, which are computed across 36 sub-boxes in the test dataset.} 
    \begin{tabular}{c|c|c|c|c}
    \hline
    \hline
        field & MSE & MAE & SSIM & PSNR  \\
        \hline
        photo-$z$ & $8.82 \pm 0.62$ &$1.52 \pm 0.07$ & $0.43 \pm 0.04$& $28.67 \pm 0.96$\\
        \hline
        spec-$z$ & $33.21 \pm 2.59$ &$1.37 \pm 0.08$ & $0.52 \pm 0.02$ &$22.92 \pm 0.99$\\
        \hline
        UNet & $5.10 \pm 0.67$ &$1.01 \pm 0.05$ & $0.70 \pm 0.02$ &$31.07 \pm 0.93$\\
        \hline
    \end{tabular}
    \label{tab:mse}
 \end{table}

Tab.~\ref{tab:mse} presents the mean values of MSE, MAE, SSIM, and PSNR, along with their associated 1$\sigma$ statistical uncertainties. These metrics are calculated based on 36 sub-boxes from the test dataset. As shown, the UNet model achieves the lowest values for MSE and MAE, as well as the highest values for SSIM and PSNR, demonstrating its high accuracy in reconstructing the density fields. Furthermore, the statistical errors are at least one order of magnitude smaller than the mean values, indicating that the reconstruction process is both stable and robust. 

\begin{figure*}
    \includegraphics[width=0.45\textwidth]{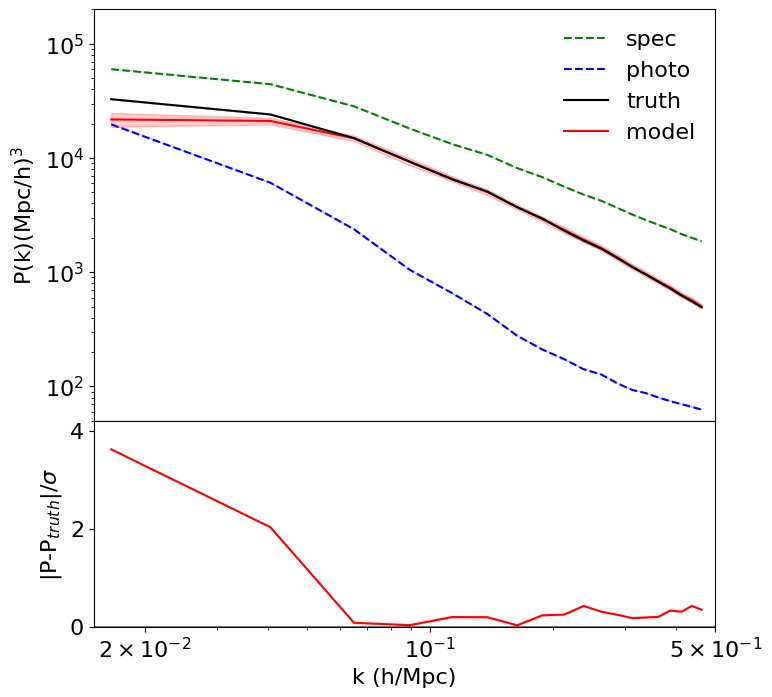}
    \includegraphics[width=0.45\textwidth]{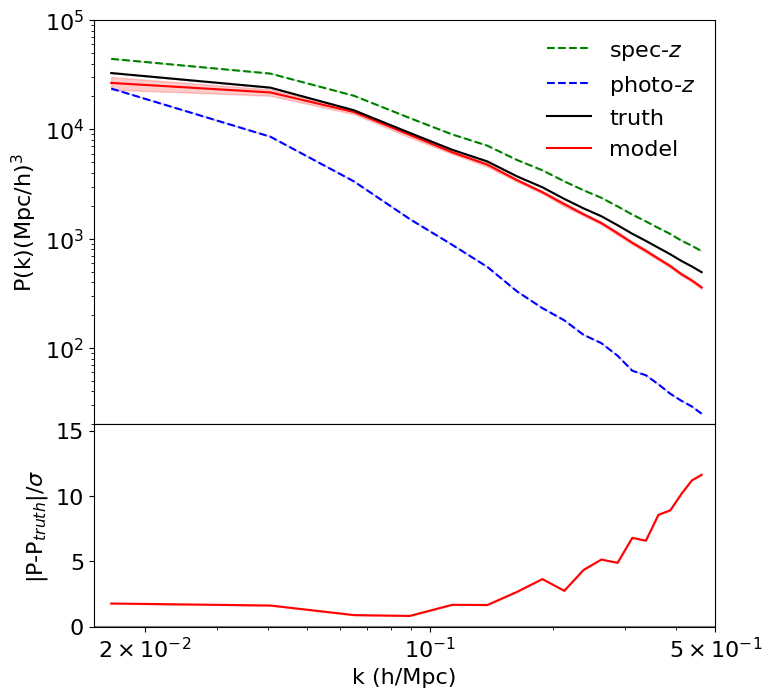}
    \caption{Comparison of the one-dimensional power spectra for the density fields of spec-$z$ (green), photo-$z$ (blue), the ground truth (black), and the UNet-predicted field (red). The left panel shows the auto power spectrum for each field, while the right panel displays the cross power spectrum between each field and the true field. In the lower panels, the relative deviations between the predictions and the ground truth, normalized to the associated $1\sigma$ statistical uncertainty, are shown. For small scales with $k \gtrsim 0.066~h/\rm Mpc$, the UNet model successfully reconstructs the amplitude of the density field, with results consistent within 0.024--0.42$\sigma$ for the auto power spectrum. The model significantly improves the accuracy of the cross power spectrum, with relative deviations reaching $0.81\sigma$ at best, suggesting that the phase can be recovered to some extent. However, for larger scales with $k \lesssim 0.066~h/\rm Mpc$, the predicted auto- and cross-power spectra are underestimated by 2.02--3.61$\sigma$ for the auto spectrum, and 1.61--1.76$\sigma$ for the cross spectrum. This discrepancy may arise from the limited information available for large-scale features when training the UNet on small boxes, which restricts its ability to effectively capture the low-$k$ Fourier modes.
 }       \label{fig:ps1d}
\end{figure*}

Fig.~\ref{fig:ps1d} presents a comparison of the one-dimensional power spectra for density fields derived from spec-$z$ (green), photo-$z$ (blue), ground truth (black), and the UNet-predicted field (red). The left panel displays the auto power spectra for each field, while the right panel shows the cross power spectra between each field and the true field. The lower panels depict the relative deviation from the 1$\sigma$ statistical uncertainty. 

As shown in the left panel, for small scales with $k \gtrsim 0.066~h/\rm Mpc$, the UNet model effectively reconstructs the amplitude of the density field, with results consistent within the $0.024$--$0.42\sigma$ range. However, for smaller values of $k$, the model tends to underestimate the auto power spectrum by more than $2\sigma$. This discrepancy is likely due to the limited information available for large-scale features during training on small boxes, which reduces the model's ability to capture low-$k$ Fourier modes effectively.

From the right panel, the reconstruction accuracy is lower than that for the auto power spectrum, primarily due to inaccuracies in the phase reconstruction of the density field. Nevertheless, the UNet model can still reconstruct the cross power spectrum for $k \lesssim 0.066~h/\rm Mpc$ within the $1.61$--$1.76\sigma$ range. As $k$ increases beyond $0.1~h/\rm Mpc$, the relative deviation gradually increases and exceeds the $2\sigma$ level.

Note that both spec-$z$ and photo-$z$ fields exhibit significant deviations from the true power spectrum. However, the UNet model effectively reconstructs the true density field by combining information from these two surveys, demonstrating its strong effectiveness in integrating spec-$z$ and photo-$z$ data for accurate reconstruction. It can also be observed that the auto and cross power spectra of the spec-$z$ samples exhibit the highest amplitudes. This is because these samples are selected based on massive halos and subhalos, which introduces a larger bias and results in the higher power spectrum amplitudes.


\begin{figure*}
    \centering
    \includegraphics[width=\textwidth]{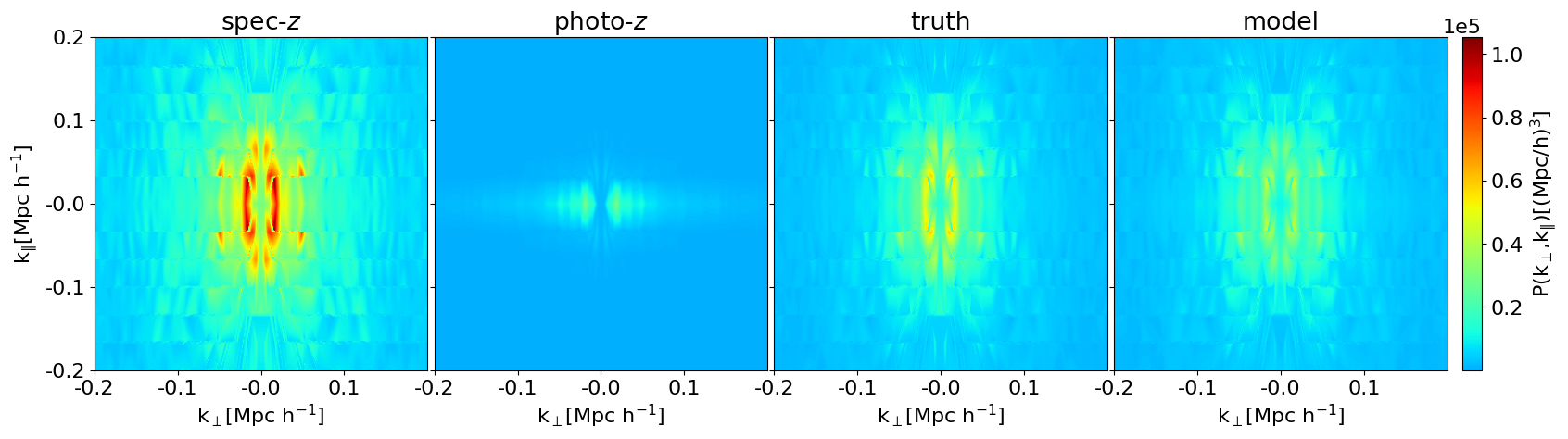}
    \caption{Comparison of the two-dimensional power spectra of density fields for the spec-$z$, the photo-$z$, the ground truth, and the reconstruction (from left to right), at $z=0.59$. The analysis is based on a slice of the large test data cube along LoS, with a side length of  $1200 \times 1200 \times 200~({\rm Mpc}/h)^3$. Specifically, the UNet model was applied to 36 smaller test boxes within this cube.
   }
    \label{fig:ps2d}
\end{figure*}

Fig.~\ref{fig:ps2d} displays the two-dimensional power spectra of the density fields for the spec-$z$, the photo-$z$, the ground truth, and the UNet reconstruction, providing a comprehensive comparison. The power spectra are computed using a slice of the test data cube at $z=0.59$, with side length of $1200 \times 1200~{\rm Mpc}/h$ in the plane perpendicular to LoS and $200~{\rm Mpc}/h$ along LoS. The the spec-$z$ and the photo-$z$ fields exhibit power spectrum patterns that significantly deviate from that of the true density field. However, the UNet model effectively reproduces a power spectrum pattern that agrees well with the ground truth, demonstrating its capability to accurately reconstruct the true density field.
\begin{figure*} 
    \centering
    \includegraphics[width=0.3\textwidth]{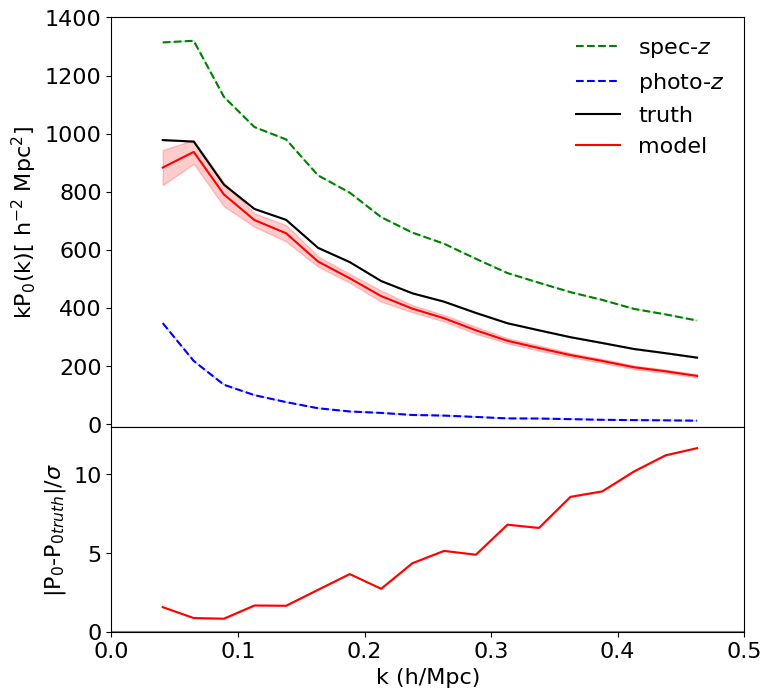}
    \includegraphics[width=0.3\textwidth]{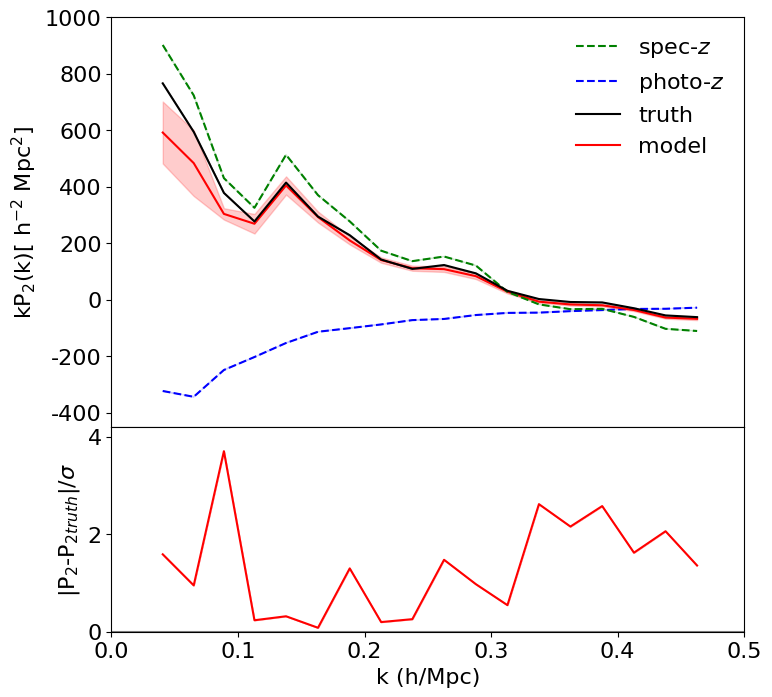}
    \includegraphics[width=0.3\textwidth]{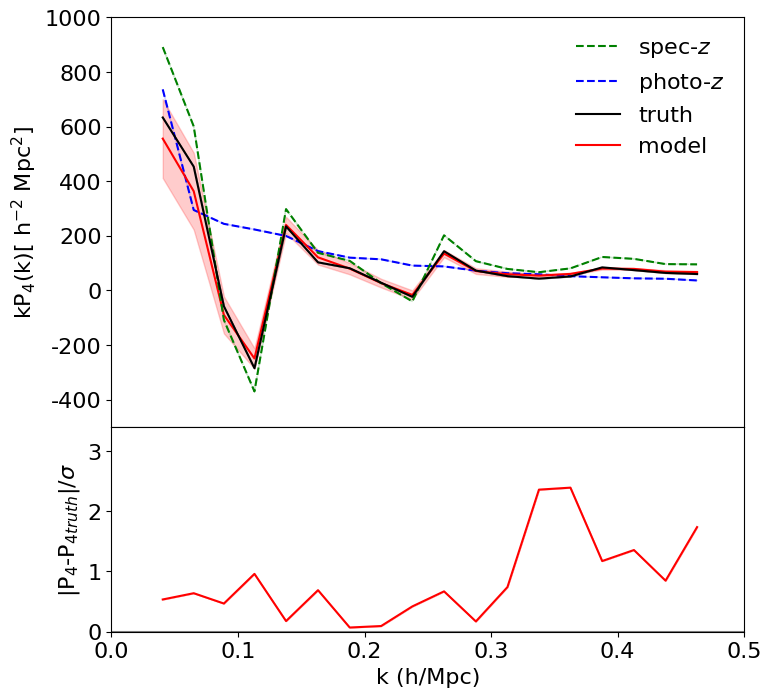}
    \caption{Power spectrum multipoles, $P_{\ell}$, for the monopole ($\ell=0$, left), quadrupole ($\ell=2$, middle), and hexadecapole ($\ell=4$, right). The $1\sigma$ error bars indicate the statistical uncertainties in the measurements, derived from the test mocks at redshift $z=0.59$. Upper panel: the power spectra for spec-$z$ (green) and photo-$z$ (blue) fields, the truth (black), and the UNet reconstruction (red) are shown for comparison. Lower panel: the relative deviations from the true values, normalized to the associated statistical uncertainties, are shown for the UNet reconstruction. Due to the significantly larger relative deviations for the spec-$z$ and photo-$z$ cases, their results are not presented.}
    \label{fig:ps_pole}
\end{figure*}
More quantitatively, the resulting power spectrum multipoles, $P_{\ell=\{0,2,4\}}$, at $z=0.59$, are shown in Fig.~\ref{fig:ps_pole}. As known, the monopole, $P_0$, represents the isotropic component of the power spectrum, while the quadrupole, $P_2$, and hexadecapole, $P_4$, capture anisotropic features primarily due to redshift-space distortions. Without the UNet reconstruction, the power spectrum multipoles for photo-$z$ deviate significantly from the true values across all scales. Although the deviations for spec-$z$ are reduced, they remain noticeable. After applying the UNet reconstruction, the reconstructed multipoles generally agree with the true values. 

Specifically, for $P_0$ at $k \lesssim 0.1~h/\rm Mpc$, the reconstructed spectrum agrees well with the true values within the $2\sigma$ level. However, as $k$ increases, the monopole is progressively underestimated beyond $2\sigma$, although it is still significantly improved compared to the green (spec-$z$) and blue (photo-$z$) lines. For $P_2$, the UNet-predicted spectrum deviates from the true value by no more than $2\sigma$ for $k\in [0.1,0.3]~h/\rm Mpc$. 
Similarly, for $P_4$, the reconstructed spectrum shows slight deviations across all scales, but remains consistent with the true values within the $2\sigma$ level for $k \lesssim 0.3~h/\rm Mpc$ on average. 
Overall, the UNet performs well by integrating the information from both spec-$z$ and photo-$z$ samples, effectively mitigating the photo-$z$ errors and leading to a high-quality reconstruction of the power spectrum of the true photo-$z$ field.


\section{Conclusion and Discussion}\label{sect:con}

Galaxy surveys are essential for tackling key cosmological questions by mapping LSS of the Universe. Imaging surveys offer extensive sky coverage and valuable cosmological information but are hindered by the challenge of inaccurate photo-$z$ measurements. Spectroscopic surveys provide precise 3D clustering and accurate redshifts but are limited in the size of galaxy samples. 

In this study, the goal is to combine these two types of surveys to take advantage of their strengths while mitigating their respective limitations. We introduce a novel deep learning approach that utilizes a dual UNet neural network to integrate photo-$z$ and spec-$z$ data at the field level. This method optimizes the extraction of information from both data types, improving redshift accuracy and offering a more robust analysis of the LSS of the universe.

We assessed the performance of the dual UNet model by using it to reconstruct the true photo-$z$ density field in the absence of redshift errors. The network employs 3D convolutional layers and a Loss function based on MSE to optimize reconstruction accuracy. The model was trained using the CosmicGrowth simulations, specifically the $z = 0.59$ snapshot, which employs $2048^3$ particles within a volume of $(1200~h/\rm Mpc)^3$ under a $\Lambda$CDM cosmology. The data is split into training and testing sets, with approximately 75\% of the data used for training (108 sub-boxes) and 25\% for testing (36 sub-boxes). Additionally, a correction is applied to adjust the amplitudes in the reconstructed density field at lower $k$-modes, as the training was conducted on smaller boxes that lack low-$k$ information.

The performance of the dual UNet model was evaluated in reconstructing density fields from a test dataset consisting of 36 sub-boxes, each with a physical size of $(200~h/\rm Mpc)^3$ and containing $100^3$ pixels. The model's accuracy was evaluated using multiple performance metrics, including the correlation coefficient, mean absolute error (MAE), mean squared error (MSE), peak signal-to-noise ratio (PSNR), and structural similarity index (SSIM). The results demonstrate that the network achieves high-quality reconstruction when integrating both spec-$z$ and photo-$z$data.

For the power spectra, the dual UNet model accurately reconstructed the photo-$z$ density field at small scales ($k \gtrsim 0.1h/\rm Mpc$), within 1--2$\sigma$ of the true values, but showed reduced accuracy for smaller $k$ values, likely due to the limited training on the smaller simulation boxes. The reconstruction based on the cross power spectrum is accurate within $2\sigma$ for $k \lesssim 0.1~h/\rm Mpc$, but its accuracy decreases as $k$ increases.

For the monopole ($P_0$), at $k \lesssim 0.1h/\rm Mpc$, the reconstructed spectrum agrees well with the true values within the $2\sigma$ level. However, as $k$ increases, the monopole is progressively underestimated, with deviations exceeding $2\sigma$ at larger $k$ values, though it still shows a significant improvement over the spec-$z$ (green line) and photo-$z$ (blue line) predictions. For the hexadecapole ($P_2$), the UNet-predicted spectrum deviates from the true values by no more than $2\sigma$ for $k \in [0.1, 0.3]~h/\rm Mpc$.Similarly, for the quadrupole ($P_4$), the reconstructed spectrum shows slight deviations across all scales but remains within the $2\sigma$ range for $k \lesssim 0.3h/\rm Mpc$ on average.

Overall, the dual UNet model exhibits robust performance by effectively combining information from both spec-$z$ and photo-$z$ samples. This integration significantly reduces photo-$z$ errors, resulting in high-fidelity reconstructions of the power spectrum across a wide range of scales. In the future, further testing of the dual UNet model across a broader range of simulations and covering different cosmological parameters will be essential for assessing the generalizability and robustness of the network. More importantly, applying the model to real observational datasets will provide valuable insights into its performance in practical, real-world conditions.

\section*{Acknowledgements}
We gratefully acknowledge Prof. Yipeng Jing for providing us the CosmicGrowth cosmological simulation suite. This work is supported by the National SKA Program of China (2020SKA0110401, 2020SKA0110402, 2020SKA0110100), the National Key R\&D Program of China (2020YFC2201600), the National Science Foundation of China (12373005, 12473097), the China Manned Space Project with No. CMS-CSST-2021 (A02, A03, B01), and the Guangdong Basic and Applied Basic Research Foundation (2024A1515012309) and the Fundamental Research Funds for the Central Universities, Sun Yat-sen University(No.24qnpy122). We wish to acknowledge the Beijing Super Cloud Center (BSCC) and Beijing Beilong Super Cloud Computing Co., Ltd (http://www.blsc.cn/) for providing HPC resources that have significantly contributed to the research results presented in this paper.

\section*{Data Availability}
The \texttt{CosmicGrowth--WMAP\_2048\_1200} simulation used in this study is available  the Simulation Database (see the  \href{https://github.com/cosmologywater/SimulationsOnClusters/blob/main/cosmicGrowth.txt}{url}).

\bibliography{apssamp}

\end{document}